\documentclass[acmsmall,natbib=false]{acmart}
\usepackage{algorithm}
\usepackage{algorithmic}
\usepackage{amsmath}

\AtBeginDocument{%
  }

\setcopyright{acmlicensed}
\copyrightyear{2025}
\acmYear{2025}
\acmJournal{JACM}
\begin{document}

\title{FAIR-CS: Framework for Interdisciplinary Research Collaborations in Online Computing Programs}

%% (If you have specific authors, update their info. Otherwise, placeholders are used here.)
\author{Breanna Shi}
\affiliation{
  \institution{Georgia Institute of Technology}
  \city{Atlanta}
  \country{US}
}
\email{bshi42@gatech.com}

\author{Thomas Deatherage}
\affiliation{
  \institution{Georgia Institute of Technology}
  \city{Atlanta}
  \country{US}
}
\email{tdeatherage3@gatech.edu}
\author{Jeanette Schofield}
\affiliation{
  \institution{Georgia Institute of Technology}
  \city{Atlanta}
  \country{US}
}
\email{jeanette@gatech.edu}
\author{Charles R. Clark}
\affiliation{
  \institution{Georgia Institute of Technology}
  \city{Atlanta}
  \country{US}
}
\email{cclark339@gatech.edu}
\author{Thomas Orth}
\affiliation{
  \institution{Georgia Institute of Technology}
  \city{Atlanta}
  \country{US}
}
\email{torth8@gatech.edu}
\author{Nicholas Lytle}
\affiliation{
  \institution{Georgia Institute of Technology}
  \city{Atlanta}
  \country{US}
}
\email{nlytle3@gatech.edu}

\renewcommand{\shortauthors}{Shi et al.}

\begin{abstract}
Research experience is crucial for computing master's students pursuing academic and scientific careers, yet online students have traditionally been excluded from these opportunities due to the physical constraints of traditional research environments. This paper presents the Framework for Accelerating Interdisciplinary Research in Computer Science (FAIR-CS), a method for achieving research goals, developing research communities, and supporting high quality mentorship in an online research environment. This method advances virtual research operations by orchestrating dynamic partnerships between master's- level researchers and academic mentors, resulting in interdisciplinary publications. We then discuss the implementation of FAIR-CS in the Human-Augmented Analytics Group (HAAG), with researchers from the Georgia Tech's Online Master of Computer Science program. Through documented project records and experiences with 72 active users, we present our lessons learned and evaluate the evolution of FAIR-CS in HAAG. This paper serves as a comprehensive resource for other institutions seeking to establish similar virtual research initiatives, demonstrating how the traditional research lab environment can be effectively replicated in the virtual space while maintaining robust collaborative relationships and supporting knowledge transfer. 
\end{abstract}


\maketitle

\section{Introduction}
The landscape of graduate computing education has undergone a fundamental transformation with the emergence of large-scale online programs, dramatically improving accessibility and affordability of advanced degrees \cite{park2020affordable, kumar2017mentoring, ross2017campus}. Research indicates that online computing programs particularly appeal to intrinsically motivated learners, notably older professionals and women seeking career transitions. These programs enable working professionals, who form the majority of enrollments, to pursue advanced education while maintaining their careers \cite{duncan2020}. While these programs address many limitations of traditional in-person programs, questions remain about how effectively these programs address the unique needs and aspirations of their student population. Creating an equitable graduate experience for these historically under-served communities presents challenges that require collaborative solutions.\\
This educational evolution has exposed a gap in research opportunities between online and in-persons masters programs which could put online masters students at a disadvantage applying for doctoral programs as admissions often give weight to prior research experience. While institutions have successfully adapted course delivery for online audiences, the translation of research experience has not kept pace \cite{brusilovsky2020}. Online research introduces substantial costs for both students and supervisors, from investments in home computing infrastructure to communication demands that frequently extend beyond traditional working hours \cite{nasiri2015postgraduate}. Advancing online research requires institutions to prioritize solutions for community building, establish clear expectations, ensure technical competency, and provide robust support for faculty mentors \cite{pollard2021mentoring}. With greater incentives for faculty participation in online-mentorship, we could see greater participation in research to match that of in-person students. \\
Running parallel to this demand in opportunities for computing research is a rising demand for interdisciplinary research - for collaboration between domain scientists and computer scientists \cite{Kirsch2020BreakingDT, Fagan2023ImmunologyAP, Salmela2021InternallyII}. However, computational faculty often limit their publications to their primary field rather than sharing findings across disciplines, impeding potential interdisciplinary partnerships. This tendency largely stems from academic departments' traditional structure and promotion systems that reward specialized contributions within singular disciplines over interdisciplinary work \cite{Yang2024DoesII, Mkinen2024InterdisciplinaryRT}. Recent advances in computing techniques can be readily applied and tested in interdisciplinary settings allowing for advancement in the domain subject area while simultaneously giving students seeking an understanding of these techniques an applied setting to learn new skills. Interdisciplinary projects, while challenging to staff through conventional means, present ideal opportunities for online computational researchers when supported by collaborative mentorship structures.\\
While research groups seeking to expand into interdisciplinary computing face persistent difficulties in recruiting computational talent, online master's students simultaneously remain under-served in research opportunities, primarily due to infrastructure limitations and competition with in-person students. This paper presents a solution that leverages a complementary opportunity at the intersection of computer science education and  interdisciplinary research. We advance this solution through four key components: (1) a comprehensive examination of current practices in online mentorship and interdisciplinary programming, (2) the introduction of our Framework for Accelerating Interdisciplinary Research in Computer Science (FAIR-CS), (3) a implementation analysis of FAIR-CS within the Human-Augmented Analytics Group (HAAG), and (4) evidence-based recommendations for future practitioners implementing the FAIR-CS framework. 

\section{Previous Work}
The effective translation of research experiences into online environments requires careful consideration of established mentoring practices and their potential adaptation to digital contexts. Qualities for successful mentoring include demonstrating genuine care for students, maintaining accessibility, serving as professional and personal role models, providing individualized guidance, and actively facilitating students' integration into the professional community \cite{bloom2007}.
 The scaling of online research programs presents challenges previously encountered in large-scale online education \cite{joyner2020global, shah2020global, joyner2017scaling, joyner2020peripheral}. Strategies that emphasize the development of comprehensive shared
resources and the implementation of structured peer mentoring systems have emerged and maintain positive student experience while accommodating larger student cohorts \cite{hudgins2020, duncan2020}. These student support solutions can be adapted for online research mentorship.

While extensive research exists on traditional graduate student mentoring practices, the effective implementation of mentoring in online environments remains an understudied area. A review conducted by Pollard and Kumar \cite{pollard2021mentoring} identified challenges including communication barriers due to reduced social presence, the absence of non-verbal cues, and limitations of asynchronous interaction, particularly in globally distributed virtual labs. However, research also highlights distinctive advantages of online mentoring, including enhanced flexibility, comprehensive documentation of communications and assignments, and the potential for scaled research operations beyond traditional lab capacities \cite{duffy2018, kumar2017mentoring, kumar2019dissertations, lechuga2011faculty, ross2017campus}. The critical need for institutional support and structured training programs for faculty, as many professors lack formal preparation in managing online research teams and mentoring relationships. Notably, frequent communication has been identified as essential for successful online mentoring relationships \cite{broome2011, jacobs2015, kumar2017doctoral}, helping to foster immediacy and reduce temporal distance that can impede effective mentorship \cite{duffy2018, nasiri2015postgraduate}. While further research is needed to explore strategies, online environment design, program structure, and supportive elements that benefit online mentees, these students maintain high autonomy by continuously negotiating their needs for structure, dialogue, and support throughout each stage of the research process \cite{kumar2017mentoring}. These deliberate designs are implemented in FAIR-CS.

\section{Framework for Accelerating Interdisciplinary Research in Computer Science (FAIR-CS)}

\begin{figure}[h]
  \centering
  \includegraphics[width=\linewidth]{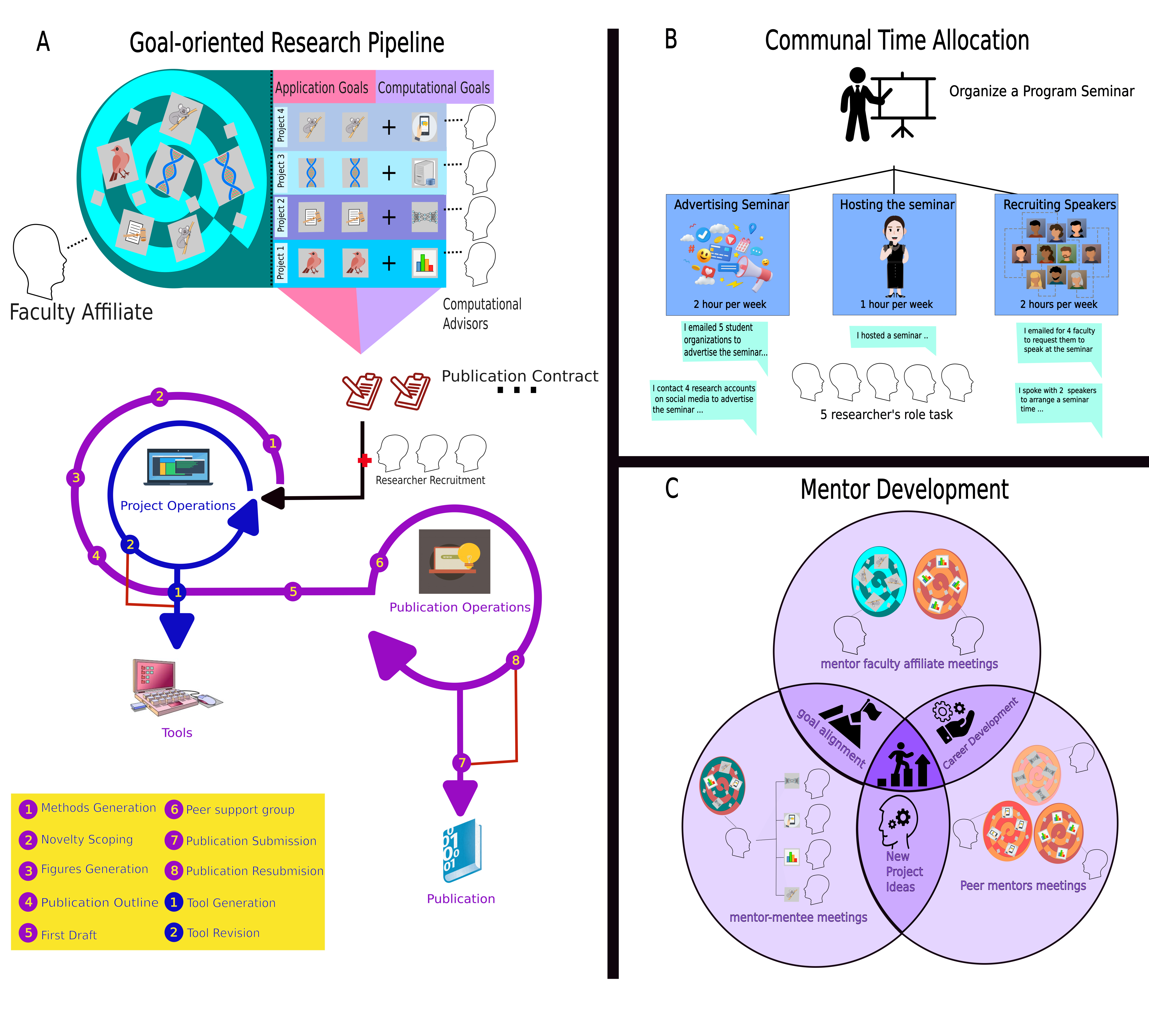}
  \caption{An overview of the FAIR-CS method. A. Goal-oriented Research Pipeline, B. Communal Time Allocation, C. Mentor Development}
  \Description{FAIR-CS Figure}
  \label{fig:FAIRsystem}
\end{figure}

The Framework for Accelerating Interdisciplinary Research in Computer Science (FAIR-CS) is a practical management structure that helps cross-disciplined research teams effectively collaborate to achieve specialized applied computing research goals (Figure \ref{fig:FAIRsystem}). FAIR-CS is composed of 3 core principles (1) goal-oriented research pipeline, (2) communal time allocation, and (3) mentor development. The interdisciplinary research process is inherently exploratory, so research operations practices must be flexible to allow contributors to shape the process of the project at various stages. However, the logistics required to facilitate cross-discipline computing research can be burdensome without an organized engagement structure. We will present how FAIR-CS can help tackle the inherent challenges presented in interdisciplinary research at scale; first, we define the necessary roles: 

\begin{enumerate}
\item \textbf{Faculty affiliates} -- Interdisciplinary faculty with application-based expertise and variable computational expertise who need computational talent to develop an application-based tool. The faculty affiliate's role is to advise (or to assign a member of their research group to advise) on all aspects of the project related to the application. The faculty affiliates will assess the tool based on its fitness to benefit a broader community interested in the faculty’s area of expertise. In addition, they must provide all data necessary for computational research.

\item \textbf{Computational advisors} -- Volunteer-based PhD or Postdoctoral researchers in computational or quantitative fields. The computational advisors serve as the primary mentors for the researchers; they utilize their expertise to advise researchers in best practices for generating the tool requested by a faculty affiliate and further advise on packaging the completed work into a format suitable for an applied computing publication.  Computational advisors receive authorship credit for publications they advise.    

\item \textbf{Researchers} -- Computational students enrolled in a credit-based research course and assigned to a project. Their primary role is to perform the requested computation and documentation tasks to satisfy a shared tool generation \& publication outcome.
\end{enumerate}

The roles define a division of labor between mentees and mentors. It is notable that the role of mentorship is split between application and computational goals, so that each mentor trains their mentee in their own expertise, allowing the mentee to perform cross-disciplinary tasks. This paper outlines best practices for collaboration among these three roles and the administrative processes that can enable a research program using FAIR-CS principles to grow.

\subsection{Goal-Oriented Research Pipeline}
The FAIR-CS research pipeline dictates how the researchers, computational advisors, and faculty affiliates are to build the tool the faculty affiliate requested and write the subsequent publication. As shown in Figure \ref{fig:FAIRsystem}A, the pipeline contains the following steps: (1) the synthesis of a publication contract, (2) researcher recruitment, (3) project operations, and (4) publication operations. 

\subsection{Publication Contract}
The method begins with the faculty affiliate meeting with a computational advisor to create a publication contract. Together, this pair will determine a set of application and computational goals where both agree that the completion of which would be substantial enough to warrant writing a scientific publication. Table \ref{table:PubContractTypes} outlines the scope of goals to be presented by the two advisor types. 

\begin{table}
\caption{Publication Contract Types}
\begin{tabular}{l p{10cm}}
\toprule
Goal Type & Outcome Expectations \\
\midrule
Application-Based & 
1. Metrics of significance to extract from the provided data.\\[0.5ex]
& 2. User-interface specifications for the purpose of the research product's use by a non-computational stakeholder.\\[0.5ex]
& 3. Automation of a previously manual application-based task. \\
\midrule
Computational-Based & 
1. Integration of a novel method.\\[0.5ex]
& 2. Contributions to overall performance that surpass previous works.\\[0.5ex]
& 3. Development of a novel benchmark. \\
\bottomrule
\label{table:PubContractTypes}
\end{tabular}
\end{table}

These research goals' documentation becomes the publication contract. The advisors will define research goals, the necessary skills for the project,the number of researchers requested, and provide all necessary project background.

\subsection{Researcher Recruitment}
The faculty affiliate and the computational advisor determine the number of researchers needed for the project and the skills they should have. The computational advisors advertise the publication contract through appropriate channels and select candidates by an assessment of their skills and effort levels (an example of such an assessment is given in Section \ref{sec:HAAGRecruitment}).  The computational advisors confirm their selection with the faculty affiliate.  

\section{Project Operations}
Project operations include all tasks the researchers complete to build the tool outlined in the research goals. Researchers are expected to independently learn all the necessary technical skills; the level of independence will vary by skills and effort levels. Each researcher is responsible for contributing equally and regularly meeting goals the computational advisor set. Each researcher's progress is tracked through weekly ‘proof of work’ document submissions which can contain videos or images along with a text description of the work they completed. Computational advisors review these documents and provide feedback on the completeness of the goals.  

\subsection{Publication Operations }
Described in Table \ref{table:PublicationDocs}, researchers are to complete a series of publication-based documents on the following topics: methods, project scope, figure generation, and publication outline. These documents help them frame their publication argument, identify an appropriate submission venue, and compose their first draft by the end of an academic semester. Teams complete these documents jointly in a specified order. The document is submitted to the computational advisors who review it, give feedback, and  either pass or fail the entire document. If the document fails, the researchers need to review the feedback and resubmit before proceeding to the next document in the series.

\begin{table}
  \caption{Publication Documents}
  \label{tab:freq}
  \begin{tabular}{ll}
    \toprule
       Document Type&Learning Objectives\\
    \midrule
       Methods Generation&1. Generate a step-by-step procedure to complete project’s goals\\
       &2. Justify the steps take in the procedures using relevant literature\\
    \midrule
       Novelty Scoping&1. Understand value the methods add to the scientific community\\
       &2. Construct an abstract that showcases novelty and generalizability \\
       &\ \ \ \ of the project method\\
       &3. Identify a conference/journal suitable for your method and impact\\
    \midrule
       Figures Generation&1. Define the most impactful aspects of your project related\\
       &\ \ \ \ to the dataset, the methodology, and the results.\\
       &2. Use illustrative tools to represent the impactful aspects you defined\\
    \midrule
       Publication Outline&1. Use previous literature from your selected conference/journal to \\
       &\ \ \ \ finalize the sections of your publication.\\
       &2. Outline the points to be articulated in each section of your paper\\
       
\bottomrule
\label{table:PublicationDocs}
\end{tabular}
\end{table}

\subsection{Peer Support Group }
When a project reaches a successful first publication draft, they move into an iterative cycle of revisions which persists until it is successfully published. The researchers at this stage will have an additional weekly meeting with the administrative team that publishing researchers attend to receive advice on their current draft and learn best practices in publishing from other publishing/published researchers' trial and error. This applies to researchers who are at the first draft and beyond. Topics this support group covers include journal/conference competitiveness, previous submissions reviewers' feedback, and review process expectations.

\subsection{Publication Submission }
Prior to publication submission, the publication will receive at least one full review by the program director, faculty affiliate, team leader, and computational advisor. In addition, the team support researcher performs a technical review where they read the final publication to identify violations of the journal/conference guidelines. These expectations outline the primary training in FAIR-CS is related to publication practices. The expectation is that the publication's technical problems will be primarily solved by the researchers individually, with occasional assistance from team members and their computational advisors. FAIR-CS does not discern between researchers' levels of expertise. Instead it is assumed that all members can contribute equally; the time it takes the researcher to do so may differ. The publication's progress is a results-based expectation.  

\section{Communal Time Allocation }
FAIR-CS requires all researchers to contribute to the program's operations. The primary method to do this is by segmenting and then categorizing program tasks into repeatable role tasks and delegating to the researchers, who must contribute one hour toward them weekly. Role tasks have been used to run seminars (see Figure \ref{fig:FAIRsystem}B), develop the program website, manage meetings, and track researcher submissions. The role tasks are designed to create a community and facilitate cross-team collaboration without over-burdening the mentors. The one-hour contributions are also small enough to not distract researchers from their primary goals.  

Another aspect of the collaboration effort is their project documentation. Researchers are expected to keep documentation of all submission, reviews, meetings, and code bases. To do this, FAIR-CS uses a ‘glass-house’ approach: all meetings are recorded and uploaded to the team page on the main program website. Reports, slides, and GitHub repositories are also linked to the team website. This helps the program with on-boarding new members for all role-types. Projects with privacy concerns should have the accessibility of the documentation adjusted as needed.  

\section{Mentor Development }
A researcher’s primary mentor is their computational advisor. As demonstrated in Figure \ref{fig:FAIRsystem}C, the FAIR-CS model enables inter-role relationships to aide the computational mentors' development.

\textbf{Faculty Relations}

The computational advisors will meet with the faculty affiliate associated with their work at least once a month. The computational advisors should set a topic for this meeting beforehand. Some relevant topics include career development in academia, interdisciplinary grant proposals, and ideas for future projects. The computational advisor should also prepare a report on the status of the mentee’s contributions and ensure that current tasks are aligned with the faculty expected outcomes. The computational advisor should use this time to seek guidance on any application-related information they need. The computational advisor should leave these conversations with a better understanding that will assist help them lead their researchers.  

\textbf{Mentor to Mentor }

The computational advisors meet with their peer computational advisors at least once every 2 weeks. Here, computational advisors should provide updates on their researchers' progress and share successes/failures for other mentors to learn from. Some topics discussed may include best practices for assessing the researchers' work, best uses of advising time, frequent questions or concerns among the researchers, technical tools that researchers found useful, and computational theory or literature that may assist a project. The computational advisors should leave these conversations with practical information to deploy in their leadership with their researchers.  

\textbf{Mentor to Mentee}

Computational advisors are expected to meet with their researchers for at least an hour every 2 weeks. Explicit research goals are set by the computational advisors at the end of each meeting, where researchers are expected to present their work from the previous weeks. The computational advisor is expected to manage the structure of the meeting. Some tasks for the meeting include investigating the researchers' completed work, learning the research ambitions for the next stage in the project, supporting the researchers' theoretical development through literature suggestions, and setting the researchers' goals to be completed prior to the next meeting.  

\section{Human-Augmented Analytic Group (HAAG)}
The Human Augmented Analytics Group (HAAG) is an interdisciplinary research program that implements the FAIR-CS method with researchers from the Georgia Tech Online Masters of Computer Science program (OMSCS). The OMSCS program has been running for over a decade with the goal of making a renowned, accredited university degree program both affordable and accessible (i.e. around \$7,000 for the full degree).  The program has over 13,000 students and over 50 courses in a variety of topics. In comparison to the on-campus program, students in the online program tend to be older—the median age is in the 30s—and most of them are employed full-time already \cite{joyner2017scaling}. 

The major challenge for scaling OMSCS research opportunities is finding qualified mentors in the form of faculty, staff, and senior PhD students at Georgia Tech. A radical rethinking of how to organize, design and lead graduate research projects, such as FAIR-CS, is necessary to give this large student base an opportunity to engage in a critical element of a graduate education and prepare a more diverse pool of students for further research opportunities. Here we show that FAIR-CS allows the HAAG to scale their available projects and faculty collaboration.

\subsection{FAIR-CS Implementation}
In the implementation of FAIR-CS, we introduce additional administrative roles which assist in the facilitation of the HAAG. In addition to the contributors, for use of the method at scale, we use 3 administrative role types: Program Director, Team Leader, and Support Researcher. The primary differences between the three roles being the number of projects that they manage. In our program, we operate with one program director, a minimum of one team leader for every 2 faculty affiliates, a minimum of one support researcher for every faculty affiliate. The support researchers report to the team leaders. The team leaders report to the program director. The ideal candidates for such administrative roles are Computational PhD students and master's students who would benefit from training in the leadership and program management responsibilities of these roles.  

The HAAG implementation of FAIR-CS has the following organizational levels: 
\begin{itemize}
    \item \textbf{Program} -- Refers to operations that span all projects and roles, such activities include community efforts (e.g., seminars), program policies, and grading expectations. The primary responsibility for the facilitation of these activities is the administrative roles.

    \item \textbf{Team} -- Refers to all operations relating to one faculty affiliate, such activities including advising meetings and faculty reporting preferences. A team will have a team leader and a support researcher to serve the specific functions of that group.

    \item \textbf{Project} -- Refers to all operations relating to one publication contract, such activities include meetings with computational advisors, review of a publication, and most weekly tasks performed by the researcher.
\end{itemize}

In this model, The faculty affiliates can host any number of projects. Each project will be assigned a computational advisor leading their project-level group towards the goals specified in the publication contract. The expansion of the program requires flexibility in FAIR-CS to allow new and returning members in all role-types at the beginning of each academic semester. For this reason, specifying expectations and responsibilities across the program assists in standardizing the outcomes and experience of participants. 
\subsection{HAAG Recruitment}
\label{sec:HAAGRecruitment}
The faculty affiliates are recruited by the program director and paired with computational advisors. The primary method of faculty recruitment has been through faculty recommendation of the program. The computational advisors are recruited based on fitness of their skill-set identified through review of their previous publications. Both the faculty and computational advisors can be affiliated with any research university. HAAG Publication Contracts are identical to what is described in the FAIR-CS method. In HAAG, the administrative team assists the computational advisors and faculty affiliates with their researcher recruitment process. When all publication contracts have been collected by the administrative team, the contracts will be advertised to the OMSCS Program through online forums. For continuing projects recruiting new researchers, additional background on the completed work will be provided in the form of the project website (see Figure \ref{fig:FAIRsystem}). Interested applicants will be asked to complete a formal application indicating their contracts of interest, professional experience, and academic coursework. In addition, each applicant will be asked to submit the following documents: 

\begin{itemize}
    \item \textbf{Research Plan} -- a week-by-week plan of semester research activities with resources stated to accomplish these tasks and contingency plan for adapting to plan failures. Applicants are advised to complete one research plan for their top-choice contract which will be used for their consideration in all projects of interest.  

    \item \textbf{Research Philosophy} -- Statement of the research feelings towards collaboration with team members and their expectations of mentors and administrative team.

\end{itemize}

\begin{figure}
  \centering
  \includegraphics[width=\linewidth]{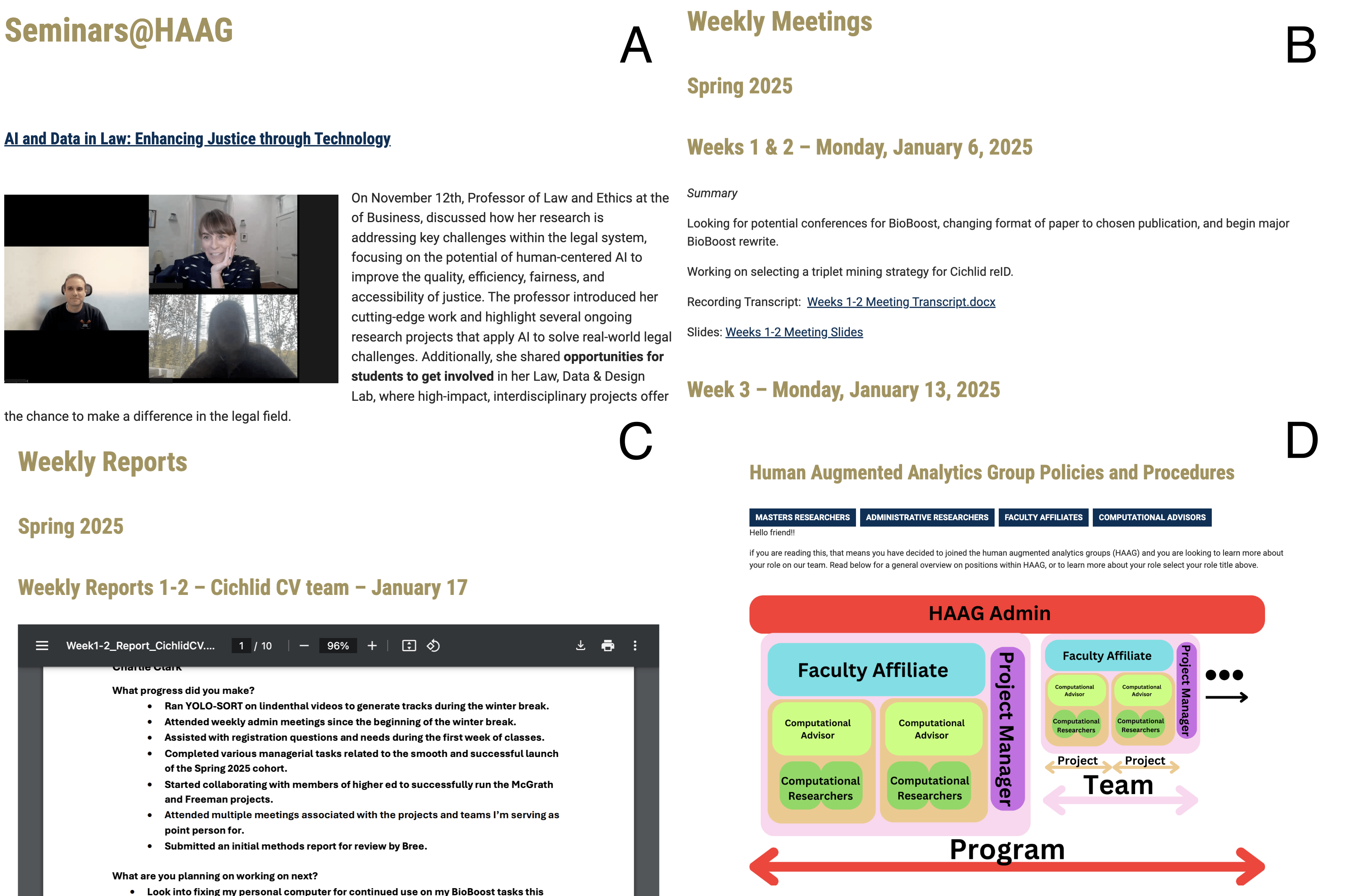}
  \caption{An overview of HAAG resource development. A. Weekly seminars B. Weekly meetings C. Weekly reports D. Research policies and procedures}
  \Description{HAAG Resource Development}
  \label{fig:HAAGResourceDevelopment}
\end{figure}

To properly match researchers to publication contracts, applicants are evaluated along four dimensions: planning detail, technical ability, technical knowledge, and willingness to collaborate. Given the nature of research, an unwillingness to collaborate is cause for immediate elimination from candidacy.  Although very few applicants are eliminated this way, it does signal to candidates that they will be evaluated in a group-collaborative context.

The level of planning detail an applicant provides is a primary factor of consideration. Similarly, technical ability is assessed by how well the applicant's research plan demonstrates ability to independently solve technical issues that may arise in the process of pursuing the contract objectives. Technical knowledge is also evaluated by considering to what extent the applicant's research plan demonstrates a familiarity with tools and concepts that will serve as resources during the completion of the contract. Planning detail, technical ability and technical knowledge are all evaluated on a 10-point scale and candidates are ranked accordingly. 

The top-ranked candidates are offered a spot for every available contract slot.  Those that accept are matched according to contracts by the following criteria: contract needs (e.g., skills needed), technical difficulty, applicants' stated preferences and team synergy, which is a holistic appraisal of the balance of skills on a contract.  As some candidates choose to decline offers, candidates are pulled off the waitlist to match the needed skills of contracts with available slots. In practice, this process works well for large applicant pools with technically skilled graduate populations. The practitioner is advised to simplify the process when working with small or novice applicant pools. Note, the support researcher roles are also advertised and selected through this mechanism. 
\subsection{HAAG Resource Development}

The HAAG implementation of FAIR-CS promotes on-going collaboration and researcher development through a number of initiatives. The HAAG website visualized in Figure 2 illustrates the resources generated by the program. Figure 2A shows a public event that are advertised beyond members of HAAG and feature speakers the broader academic community.  

Additionally, Researchers are expected to record and share notes for all their meetings (Figure 2B). All student researchers must produce weekly reports detailing their contributions, blockers, and immediate next-steps (Figure 2C). These initiatives, within the broader HAAG policies and procedures guidelines (Figure 2D), promote the on-going development of research and researcher education within a highly collaborative and open environment.

\subsection{Outcomes}
The central outcome of a project using FAIR-CS is the interdisciplinary publication. This goal allows the student to complete the full publication experience highlighting the additional skills necessary to start with an idea, generate robust results, and communicate the findings in a format suitable for a scientific audience. The computational advisor learns to manage a group of researchers independently and provide them with research goals that guides them towards publication. It notable that the FAIR-CS method differs from other mentorship models in the level of independence that PhD and Post-doc level researchers are given as advisors on the interdisciplinary project. The faculty affiliate allows the computational advisor to lead the project with the understanding that the application goals will be prioritized. For the faculty affiliate, the work can serve as proof-of-concept for grants, and the tool generated may serve a greater purpose in studies across the application discipline. The collaboration element of HAAG allows for career path outcomes. Master’s research is extremely important to a successful PhD application. This program has allowed online master’s students to be competitive for PhD applications in interdisciplinary fields such as bioinformatics and human-computer interaction. 

\section{Recommendations}
Successful interactions in a virtual research environment are not a given. FAIR-CS utilizes an intentional interaction guideline to regulate engagement. For researcher procedures, general channels are used to send information to all researchers and confirmations are requested in the form of replies to track when the information has been received by researchers. For computational advisors and faculty affiliates, it is best practice to communicate early and allow feedback when implementing procedural changes. The mentors should receive first notice when procedures may affect their operations. \\ 
Faculty affiliates set their terms of engagement for the researchers. This can involve setting up access for platforms that are used in the faculty’s research group and giving instructions for the frequency and method by which the researcher should send inquires to faculty. Faculty affiliates commonly meet with their researchers bi-weekly on the week that they are not meeting with the computational advisors, but this is not a requirement. Flexibility towards time-commitments imposed on the faculty affiliates will lead to greater growth and program satisfaction. The faculty are required to meet monthly with all computational advisors advising projects on their team to foster interdisciplinary relationships, generate new project ideas, and provide career advice to advisors interested in academic careers. \\
The computational advisors host meetings with the researchers for 1 hour bi-weekly. They also review the all documents sent to the mentor within 2-3 days of submission or within 1 week with notice of delay. Outside of their required contributions, the computational advisor may set up methods for additional interactions with the researchers. Researchers are informed that technical inquiries are to be minimal unless otherwise specified by the computational advisor to preserve the time of the mentor whose role is volunteer based. Note that such explicit control of communication could be cumbersome in an in-person setting and practitioners should make note to loosen the requirements as natural research relationships develop across the cohort.\\
The administrative team can use the scale of the projects in small conflict resolution. For example, if a researcher is struggling on a project the administration can review needs across the program to reassign members as needed. The administrative team should periodically converse with computational advisors and faculty affiliates to access researcher performance. Early-intervention leads to higher satisfaction for both the researcher and the mentors involved.
\section{Future Work}
Managing a distributed network of researchers is a challenge. The HAAG implementation of FAIR-CS relies on every single researcher dedicating an hour per week to their assigned role task, e.g., managing meetings, seminars, tracking reports, etc. Furthermore, the HAAG  implementation depends on a cadre of support researchers whose sole task is facilitating the operations of the group.  Likewise, above the support researchers are the team leaders who in turn report to the program director.  Such real-world, hierarchical project management and leadership is an excellent pedagogical opportunity for graduate level business and management students.  Moreover, as a rule, support researchers—who in HAAG are master’s computer science students--desire computational research experience rather than project management work.   Future work in FAIR-CS and HAAG should incorporate for-credit or internship opportunities for MBA students to get hands-on, real-world project management experience in an interdisciplinary context. 

\section{Conclusion}
Research experience is highly sought after among non-traditional, online computer science students. Such experience is often a critical component in career discernment, i.e., deciding to pursue a PhD or career in academia.  Such opportunities have tended to be limited to in-person academic programs. The FAIR-CS model represents one way of facilitating collaboration that mutually benefits online computer science students and faculty in an interdisciplinary context.  Moreover, we have shown the practical application of the FAIR-CS model through its implementation in the Human Augmented Analytics Group (HAAG). We encourage the adoption of FAIR-CS as an opportunity for greater collaboration in interdisciplinary research with online researchers. 

%%
%% Acknowledgments (optional)
%%
\begin{acks}
We thank all participants and mentors in the Human-Augmented Analytics Group (HAAG).
\end{acks}

%%
%% Print the bibliography
%%
%%\bibliographystyle{ACM-Reference-Format}
%%\bibliography{bibliography}
\bibliographystyle{ACM-Reference-Format}
\bibliography{bibliography}

\end{document}